\documentclass[graybox, nosecnum, natbib]{svmult}

\usepackage{mathptmx}       
\usepackage{helvet}         
\usepackage{courier}        
\usepackage{type1cm}        
\usepackage{makeidx}         
\usepackage{graphicx}        
\usepackage{multicol}        
\usepackage[bottom]{footmisc}
\usepackage{soul}            

\usepackage{aps-bibstyle}

\usepackage{bbm}
\usepackage{mathtools}
\usepackage{amsmath}
\usepackage{amssymb}

\newcommand{\boxedeqn}[1]{%
  \[\fbox{%
      \addtolength{\linewidth}{-2\fboxsep}%
      \addtolength{\linewidth}{-2\fboxrule}%
      \begin{minipage}{\linewidth}%
      \begin{equation}#1\end{equation}%
      \end{minipage}%
    }\]%
}

\newcommand{\boxedeqnlarge}[1]{%
  \[\fbox{%
      \addtolength{\linewidth}{-2\fboxsep}%
      \addtolength{\linewidth}{-2\fboxrule}%
      \begin{minipage}{1.05\linewidth}%
      \begin{equation}#1\end{equation}%
      \end{minipage}%
    }\]%
}

\renewcommand{\hslash}{\mathchar'26\mkern-9mu h}

\makeindex             

\begin{document}
\title*{Theoretical Methods for Giant Resonances}
\author{Gianluca Col\`o}
\institute{Gianluca Col\`o \at Dipartimento di Fisica, Universit\`a degli Studi di Milano, and
INFN, Sezione di Milano, 
Via Celoria 16, 20133 Milano (Italy), \email{gianluca.colo@mi.infn.it}
}
%
%
\maketitle
\abstract{
The Random Phase Approximation (RPA) and its variations and extensions are, without 
any doubt, the most widely used tools to describe Giant Resonances within a microscopic 
theory. In this chapter, 
we will start by discussing how RPA comes out naturally, if one seeks a state with 
a harmonic 
time dependence in the space of one particle-one hole excitations on top 
of the ground state. 
It will be also shown that RPA is the simplest approach in which 
a ``collective'' state emerges. 
These are basic arguments that appear in other textbooks but are also unavoidable 
as a starting 
point for further discussions. In the rest of the chapter we will give emphasis 
to developments 
that have taken place in the last decades: alternatives to RPA like the 
Finite Amplitude Method 
(FAM), state-of-the-art calculations with well-established Energy Density 
Functionals (EDFs), 
and progress in {\em ab initio} calculations. We will 
discuss extensions of RPA using as a red thread 
the various enlargements of the one particle-one hole model space. The importance 
of the continuum, 
and the exclusive observables like the decay products of Giant Resonances, 
will be also touched 
upon.
}

\section{Introduction}

Giant resonances are among the strong evidences of the fact that nuclei display
highly collective motion. Collective rotations are also among these evidences but they
take place on a slower timescale. Cluster radioactivity, or spontanoeus and induced
fission, are even slower phenomena: they go often under the name of ``large amplitude'' 
phenomena as the nucleus undergoes large fluctuations and changes in its 
configuration (\cite{Nakatsukasa2016}). On the other hand, Giant Resonances 
are ``small amplitude'' phenomena, namely small oscillations around the ground-state
configuration that are, to a first approximation, amenable to a harmonic description. 
Thus, we note in passing the remarkable fact that coherent behaviours of many, if not all, 
nucleons display at different energy scales.

Giant resonances are prominent peaks that show up in the nuclear response around
$\approx$ 10-30 MeV. ``Response'' is here a generic wording that may refer to various
external fields. If we consider a generic inelastic reaction, we can safely
state that giant resonances have a cross section which is larger than those
associated with the other states. We shall use the notation $F$ for a generic external
field and use the standard definition of strength function $S(E)$ associated
with this field, that is
\begin{equation}\label{eq:strength}
S(E) = \sum_n \left\vert \langle n \vert F \vert 0 \rangle \right\vert^2 \delta\left( 
E-E_n \right),
\end{equation}
where $\vert 0\rangle$ and $\vert n\rangle$ are the ground state and excited states,
respectively. $E_n$ is the excitation energy of the state $\vert n\rangle$.

Eq. (\ref{eq:strength}) refers to discrete states but in the case at hand we are above
the separation energy and clearly in the continuum part of the nuclear spectrum. 
The equation can be easily generalised to the case in which the states have 
an intrinsic width $\Gamma_n$, e.g., in the form
\begin{equation}\label{eq:strength_gamma}
S(E) = \sum_n \left\vert \langle n \vert F \vert 0 \rangle \right\vert^2 
\frac{\Gamma_n}{\left( E-E_n \right)^2 + \frac{\Gamma_n^2}{4}}.
\end{equation}
Indeed, in this chapter, we do refer to states that have clearly the shape of a resonance, 
i.e. are characterised by a peak energy $E$ and a conspicuous width $\Gamma$, 
and definitely emerge on top of the background of the other excited states. 

This chapter is related to theoretical methods and in essentially all of them one
calculates the whole spectrum associated with states having given quantum numbers
$J^\pi$ (total angular momentum and parity). Situations in which
the prominent states emerge much less with respect to the background, in which
the states are many and the widths are large so that resonances do overlap, or
even in which no collective behaviour shows up, may be found. In other words,
giant resonances may or may not show up in the response associated with a given 
$J^\pi$, and the physical information they carry may be very relevant or not. 

Often, one carries out studies of Giant Resonances either in order to test the
nuclear Hamiltonian and extract information like the associated Equation of State (EoS), 
or to see if the theoretical method that is employed captures all the relevant 
many-body correlations. The two aspects are not fully disconnected, though. 
To give only one example here, the simplest nuclear Giant Resonance is the 
Isoscalar Giant Monopole Resonance or ISGMR (\cite{Blaizot1980,Garg2018}). 
We shall provide in the next pages a classification of Giant Resonances, and
their names will be fully clarified. Here, let us just say that the ISGMR is also
called the ``breathing mode'' as it is a mode in which the nucleus shrinks and expands. 
Intuitively, it must be related with the nuclear compressibility. In other words,
a nuclear Hamiltonian, which is able to reproduce the properties of the 
ISGMR, should also be able to reproduce the correct value of the nuclear matter 
compressibility (taking care of experimental and theoretical uncertainties). 
Obviously, this argument holds if there are good reasons to believe that the correlation
between ISGMR energy and nuclear incompressibility is obtained in a sound many-body method.

From the historical viewpoint, the first evidence of a giant resonance has been found 
in the response of the nucleus to real photons (\cite{Berman1975}). A large resonance
has been found systematically in all nuclei, at an energy of the order of
\begin{equation}\label{eq:eneIVGDR}
E_{\rm IVGDR} \approx 80\ {\rm A}^{-1/3}\ [{\rm MeV}].
\end{equation}
At such energies, the wavelength associated with the photons, $\lambda 
= \frac{hc}{E} \approx 15\ {\rm A}^{1/3}\ {\rm fm}$, is much 
larger than the nuclear size. The protons feel the effect of an electric field
that oscillates rapidly in time but is essentially uniform in space. These
conditions correspond to the well-known dipole approximation. Protons are pulled apart
from neutrons, and the strong proton-neutron attraction provides the restoring force.
This may be seen as a forced harmonic motion: one expects a resonance when the 
external frequency matches the eigenmode of the system. This is the so-called
Isovector Giant Dipole Resonance or IVGDR.

The study of Giant Resonances is a mature field of research, in which many basic facts have been
clarified quite some time ago. Monographs exist, that review the
classification of these collective modes and illustrate the experimental findings and 
theoretical methods, established up to the turn of the 20th century (\cite{Bortignon1998,Harakeh2001}). 
Consequently, the scope of this chapter is not to start in full detail from scratch.
We shall review the basic theory aimed to describe the Giant Resonances and then focus 
on recent accomplishments and open questions. The emphasis will be on calculations based
on Density Functional Theory (DFT) and on {\em ab initio} approaches in some cases.
Purely phenomenological calculations will be mentioned when necessary.

\section{The nuclear ground state and its response to an external field: time-dependent approaches}

Let us assume that we can start from a description of the nuclear ground state 
in terms of independent particles or quasi-particles (\cite{Ring1980,Bender2003}). 
The single-particle states are solutions of a set of equations of the type
\begin{equation}\label{eq:HF}
h \phi_i = \epsilon_i \phi_i,
\end{equation}
where $h$ is an effective Hartree-Fock (HF) or Kohn-Sham (KS) Hamiltonian while $\epsilon_i$ and
$\phi_i$ are the associated eigenvalues and eigenfunctions. The density of the system
is, then, given as
\begin{equation}\label{eq:density_r}
\rho(\vec r) = \sum_{i=1}^F \phi_i^*(\vec r)\phi_i(\vec r).
\end{equation}
The single-particle states are occupied up to the Fermi level $F$, and this defines the
ground state. 

From now on, we shall use the language of second quantisation, where $a$ and $a^\dagger$ 
are the usual fermionic annihilation and creation operators. The density (\ref{eq:density_r}) 
can be expressed in a generic basis labelled by $i, j \ldots$ and its matrix elements read
\begin{equation}\label{eq:density_basis}
\rho_{ij} = \langle a^\dagger_j a_i \rangle,
\end{equation}
when evaluated on a state $\vert\rangle$ of the system under study. Using the HF or KS 
basis defined by the solution of (\ref{eq:HF}), the density matrix
(\ref{eq:density_basis}) is diagonal if it is evaluated on the HF or KS ground state. Its
diagonal matrix elements are, respectively, 1 and 0 for the states below (holes) and 
above (particles) the Fermi level $F$. We shall use the standard notation in which
$h$ ($p$) labels a state which is below (above) the Fermi energy and is occupied (empty). 

Annihilating a hole and creating a particle is the simplest possible excitation and it is named
``particle-hole'', or $ph$: we can write it as $\vert ph^{-1}\rangle \equiv a^\dagger_p
a_h \vert 0 \rangle$. These $ph$ excitations can be viewed as small variations of the 
density matrix around its ground-state value. As we shall discuss, the excitations of 
the system including the Giant Resonances can be built and understood by using this 
$ph$ basis. Nuclei having closed shells, that do not display superfluidity, can be described 
in this manner. 

When pairing correlations produce a superfluid ground state, this description is no
longer valid. The eigenstates of the system are mixtures of holes and particles, 
described in terms of the Bogoliubov transformations
\begin{equation}\label{eq:Bogoliubov}
\alpha^\dagger_\alpha =\sum_i U_{i\alpha} a^\dagger_i + V_{i\alpha} a_i.
\end{equation}
The states $\alpha$ are called quasi-particle states. 
The Hartree-Fock-Bogoliubov (HFB) ground state
generalises the HF solution: the latter has been defined by $a_p \vert 0 \rangle = 
a^\dagger_h \vert 0 \rangle = 0$, whereas the former is defined by $\alpha_\alpha \vert 0 \rangle = 0$
for every $\alpha$. In other words, the HFB ground state is the quasi-particle vacuum, while the
HF ground state is the vacuum of particles and holes. 
The HFB equations, that generalise the HF or KS equations (\ref{eq:HF}), are
\begin{equation}\label{eq:HFB}
\left( \begin{array}{cc} h-\lambda & \Delta \\ -\Delta^* & -h^*+\lambda \end{array} \right) 
\left( \begin{array}{cc} U_\alpha \\ V_\alpha \end{array} \right) =
E_\alpha \left( \begin{array}{cc} U_\alpha \\ V_\alpha \end{array} \right). 
\end{equation}
Here, $\Delta$ and $\lambda$ are the pairing field and Fermi energy, respectively. 

In analogy with what we have just discussed in the HF case, the simplest excitations of the system 
are two quasi-particle excitations (we refer here to excitations that do not change the particle
number: a single quasi-particle excitation is an excited state of the $A\pm1$ system). 
The density matrix (\ref{eq:density_basis}) is not sufficient to describe these
excitations. We need the so-called pairing tensor as well, which is defined by
\begin{equation}\label{eq:pair_tensor}
\kappa_{ij} = \langle a_j a_i \rangle.
\end{equation}
In order to have a more compact notation, we shall define
\begin{equation}\label{eq:R}
{\cal R} = \left( \begin{array}{cc} \rho & \kappa \\ -\kappa^* & 1-\rho^* \end{array} \right).
\end{equation}
$\cal R$ is called generalised density matrix. From its form, we can see that
its variations include $ph$, $hh$, $hp$ and $pp$ excitations. It can be shown that
$\cal R$ commutes with the HFB Hamiltonian so that, in the basis of the aforementioned 
quasi-particle states, this matrix is also diagonal and its eigenvalues are equal to either 0 
or 1. Another basis, which will be mentioned in what follows, is the canonical basis, in which
$\rho$ is diagonal.  

Let us now assume that the nucleus is under the action of an external field $F$. The field
gives an initial boost to the nucleus so that wave functions and associated densities
are displaced from the equilibrium values. The following time evolution of the 
generalised density
is governed by the equation
\begin{equation}\label{eq:TDHFB}
i{\hslash} \frac{d}{dt} {\cal R} = \left[ {\cal H}+F, {\cal R} \right].
\end{equation}
In principle, one could solve the time-dependent HFB (TDHFB) equation 
(\ref{eq:TDHFB}) directly,
for instance in coordinate space. The initial condition must be given, and to simulate 
Giant Resonances one could assume a displacement of all particles, like the shift of 
protons with respect to neutrons that we have described in the Introduction in the case 
of the IVGDR. Actually the TDHFB equations are also apt to study large amplitude
phenomena. On the other hand, their solution cannot fully describe the features associated with 
dissipation, like the damping of GRs. The reason is that only the one-body
density $\cal R$ is involved and not the 2-body, 3-body $\ldots$ $A$-body densities. 

\begin{figure}[t]
\includegraphics[width=0.8\textwidth]{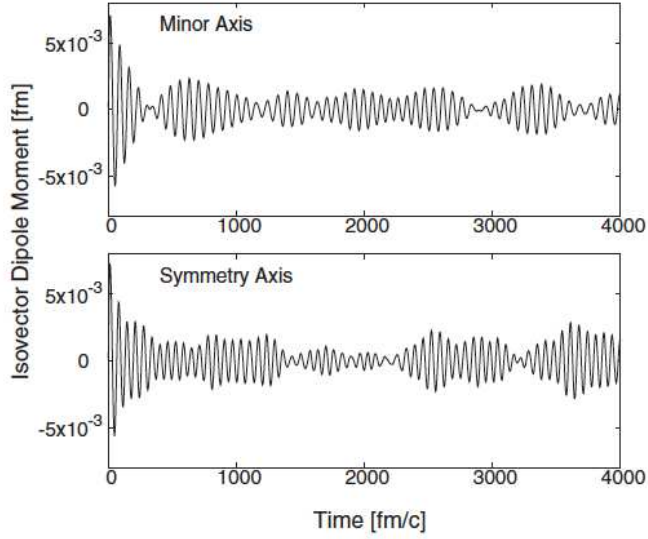}
\caption{Dipole moment as a function of time, extracted from a TDHF calculation in
the axially symmetric nucleus $^{76}$Se. Figure taken from Ref. (\cite{Goddard2013}).}
\label{fig:1}
\end{figure}

Implementations of TDHFB, or of the simpler time-dependent 
HF-Bardeen-Cooper-Schrieffer (TDHFBCS) and time-dependent HF (TDHF) 
approaches, exist. In the case of Skyrme functionals, computer codes have been 
published in the case of TDHF with only static pairing, 
and the interested reader
can look up the relevant references (\cite{Maruhn2014,Schuetrumpf2018}) 
that allow 
to grasp the details of the solution of the time-dependent equation and also to run examples 
and learn by direct experience. Other groups have also
developed similar approaches and applied them to the study of
Giant Resonances (\cite{Scamps2014}). In the past, the semiclassical 
limit of TDHF, i.e. the Vlasov equation, has been explored [see
(\cite{Baran2005}) for a review]. At the moment, this is a tool of choice
for the study of intermediate energy reactions, whereas for Giant Resonances
there is no big gain in going semiclassical. A recent comparison of
RPA, TDHF and Vlasov approaches can be found in (\cite{Burrello2019}). 
In general, there is full consistency between recent time-dependent
and RPA calculations, not only in the case of Skyrme approaches but also
in the case of covariant approaches. Relativistic time-dependent
Hartree and related approaches are well-established and 
reviewed, e.g. in (\cite{Vretenar2005}).
A computer code has been published in (\cite{Berghammer1995}).

We conclude this discussion with 
an example of a TDHF calculation for the dipole
response of a deformed nucleus, that is displayed in Fig. \ref{fig:1}. It has to be noted that the
GR strength function must be extracted from the time-dependent expectation value of the
relevant operator via a Fourier transformation. Consequently, fine details of the
strength may not be easily accessible. 

\section{Linear response: QRPA and FAM}

The further assumption is often made, that the external field $F$ is weak enough so that 
the response can be written by restricting to lowest order in the density 
variations that are encoded in $\cal R$. In practice, this is
the situation that takes place for instance in the case of direct reactions, when the
external field acts for a short time regardless of its intrinsic 
strength (or, to say it better, 
the experimental findings are often consistent with such a picture).

The weakness of the external perturbation $F$ is consistent with 
a harmonic time-dependence of the type
\begin{equation}
F(t)=F e^{-i\omega t}+F^\dagger e^{+i\omega t},
\end{equation}
where $F$ includes the dependence on space and spin. The wave functions and the
generalised density $\cal R$ (\ref{eq:R}) will also display changes with 
the same harmonic dependence:
\begin{eqnarray}\label{eq:Rt}
{\cal R}(t) & = & {\cal R}^{(0)} + \delta {\cal R}(t) \nonumber \\
\delta {\cal R}(t) & = & \delta {\cal R} e^{-i\omega t}+ \delta {\cal R}^\dagger e^{+i\omega t},
\end{eqnarray}
where ${\cal R}^{(0)}$ is the equilibrium value given by (\ref{eq:R}) evaluated in the
HFB ground state. Consequently, the Hamiltonian $\cal H$ will follow 
the same harmonic behaviour.
In these conditions, and by neglecting higher order terms
in $F$ or $\delta{\cal R}$ or $\delta{\cal H}$, one derives the so-called
quasi-particle Random Phase Approximation (QRPA) 
equations (\cite{Khan2002,Bender2003}). 

In fact, Eq. (\ref{eq:TDHFB}) becomes
\begin{eqnarray}\label{eq:small_ampli}
\hslash\omega \delta{\cal R} & = & \left[ {\cal H}^{(0)}, \delta{\cal R} \right]
+ \left[ \delta{\cal H} + F, {\cal R}^{(0)} \right], \nonumber \\
-\hslash\omega \delta{\cal R}^\dagger & = & \left[ {\cal H}^{(0)}, \delta{\cal R}^\dagger \right]
+ \left[ \delta{\cal H}^\dagger + F^\dagger, {\cal R}^{(0)} \right],
\end{eqnarray}
where the upper (lower) line refers to the terms that go like $e^{-i\omega t}$ 
($e^{+i\omega t}$). The QRPA set of equations can be obtained from the first
line only (the redundancy of the second line could be checked as an
exercise). In the above equations the term $\left[ {\cal R}^{(0)}, {\cal H}^{(0)} \right]$ 
vanishes because, as we mentioned in the previous Section, the two matrices can 
be simultaneously diagonalised in the
HFB ground-state. Finally, one can neglect the term in $F$ because the external field
plays only the role of an initial boost, as we already alluded to, and does not determine
the eigenmodes that are intrinsic properties of the system under study.

The QRPA equations are obtained by taking the matrix elements of 
the first equation in (\ref{eq:small_ampli}), either between the 
ground-state $\vert \tilde 0\rangle$ and a two quasi-particle state 
$\alpha^\dagger_\alpha\alpha^\dagger_\beta\vert \tilde 0 \rangle$, 
or between the ground-state and $\alpha_\beta\alpha_\alpha \vert \tilde 
0 \rangle$ (the ground-state $\vert \tilde 0\rangle$ is more general than 
the HFB ground state $\vert 0 \rangle$). By taking
these matrix elements, one finds the following quantities:
\begin{eqnarray}
X_{\alpha\beta} & = & \langle \tilde 0 \vert \alpha_\beta\alpha_\alpha\
\delta{\cal R} \vert \tilde 0 \rangle, \nonumber \\
Y_{\alpha\beta} & = & \langle \tilde 0 \vert \alpha^\dagger_\alpha\alpha^\dagger_\beta\
\delta{\cal R} \vert \tilde 0 \rangle,
\end{eqnarray}
that are the components of the density variation $\delta{\cal R}$ on the
two quasi-particle basis. This means that the excited states $\vert n
\rangle$ are here assumed to be  
\boxedeqn{\label{eq:QRPAstates}
\vert n \rangle = \left( \sum_{\alpha\beta} X^{(n)}_{\alpha\beta} 
\alpha^\dagger_\alpha\alpha^\dagger_\beta - Y^{(n)}_{\alpha\beta}
\alpha_\beta\alpha_\alpha \right)\ \vert \tilde 0 \rangle.
}
QRPA can be seen
as a linearisation (or a small amplitude limit) of TDHFB: only the 
amplitudes associated
with the simplest excitations, namely two quasi-particle creation and annihilation, are 
considered. We have put Eq. (\ref{eq:QRPAstates}) in a box to stress this
hypothesis, and we shall do the same with the following equations
of the same kind, so that the reader can appreciate at first sight the various
enlargements (or restrictions) of the model space. 

The matrix elements of the first line of (\ref{eq:small_ampli}) produce the QRPA
equations in their matrix form:
\begin{equation}\label{eq:matrix}
\left( \begin{array}{cc} A & B \\ -B^* & -A^*
\end{array} \right) \left( \begin{array}{c} X \\
Y \end{array} \right) = \hslash\omega \left( \begin{array}{c} X \\
Y \end{array} \right).
\end{equation}
The QRPA equations can be written in a simple form if one
employs the canonical basis, as the matrix elements become
\begin{eqnarray}\label{eq:qrpa_can}
A_{\alpha\beta,\gamma\delta} & = & \left( E_{\alpha\gamma}
+ E_{\beta\delta} \right) +
\left( u_\alpha v_\beta u_\gamma v_\delta +
v_\alpha u_\beta v_\gamma u_\delta \right)
\langle \alpha \delta \vert V \vert \beta \gamma \rangle \nonumber \\
& + & \left( u_\alpha u_\beta u_\gamma u_\delta +
v_\alpha v_\beta v_\gamma v_\delta \right)
\langle \alpha \beta \vert V \vert \gamma \delta \rangle, \\
B_{\alpha\beta,\gamma\delta} & = & 
\left( u_\alpha v_\beta u_\gamma v_\delta +
v_\alpha u_\beta v_\gamma u_\delta \right)
\langle \alpha \delta \vert V \vert \beta \gamma \rangle \nonumber \\
& - & \left( u_\alpha u_\beta v_\gamma v_\delta +
v_\alpha v_\beta u_\gamma u_\delta \right)
\langle \alpha \beta \vert V \vert \gamma \delta \rangle.
\end{eqnarray}
Here, $E$ is the matrix of the quasi-particle energies already introduced
in (\ref{eq:HFB}), when written in the canonical basis. In this basis,
such matrix is no longer diagonal. The matrices $U$ and $V$ of 
(\ref{eq:Bogoliubov}), as well as the pairing tensor (\ref{eq:pair_tensor}), 
assume a simple form in terms only of the coefficients $u$ and 
$v$ that appear in the previous equation (\cite{Ring1980}). Everything becomes
even simpler in the BCS approximation, because the quasi-particle
basis coincides with the canonical basis and $E$ is also diagonal.
Moreover, the $v_\alpha$ and $u_\alpha$ coefficients are, in this case, the 
probability amplitudes that the state $\alpha$ is occupied or empty.

Once the QRPA solution is available, with the energies 
$\hslash\omega_n$ of the states $\vert n\rangle$ and their wave functions written
in terms of the $X$ and $Y$ amplitudes, the strength function 
(\ref{eq:strength}) can be readily calculated.
More details about formulas, together with the complete derivations, 
can be found in Refs.
(\cite{Rowe1970,Ring1980,Suhonen2007}), where one also finds the 
expressions in the angular-momentum
coupled basis. In the quasi-particle basis, the matrices $U$ and
$V$ of the Bogoliubov transformation (\ref{eq:Bogoliubov}) are not diagonal, 
and this would make the calculation of the QRPA matrix 
(\ref{eq:matrix}) rather cumbersome. 

Even in canonical basis, the matrix in (\ref{eq:matrix}) may become
very large, especially for heavy and/or deformed nuclei in which spherical
symmetry cannot be assumed any longer. This fact has pushed the developments
of alternative methods to implement and solve QRPA. In recent years, 
the so-called finite-amplitude method (FAM) has become increasingly 
popular.

The FAM has been introduced in (\cite{Nakatsukasa2007}) in the case without
pairing, and extended to the case with pairing in (\cite{Avogadro2011}). The
starting point is always the time-dependent equation (\ref{eq:small_ampli})
in the linearised or small-amplitude case. The essential trick is that 
of assuming a generic small variation of $\cal R$ and calculate the 
corresponding variation of $\cal H$ in the form
\begin{equation}\label{eq:FAM}
\delta {\cal H} = \frac{{\cal H}[{\cal R}^{(0)}+\eta \delta{\cal R}] 
- {\cal H}^{(0)}}{\eta},
\end{equation}
where $\eta$ is a small parameter. It should be noted that this does not
require much more effort than a HFB calculation, as one has simply
to evaluate the generalised density and the corresponding Hamiltonian
around the ground state, as it is done when seeking for convergence of
the HFB equations. In other words, the evaluation of two-body QRPA matrix
elements (\ref{eq:qrpa_can}) is avoided. By knowing the relation between
$\cal H$ and $\cal R$ in the linear regime through Eq. (\ref{eq:FAM}), 
one has all the elements to solve Eq. (\ref{eq:small_ampli}). In the papers
we have mentioned, this is done by writing this equation on a finite
quasi-particle basis. In this way, one turns it into a linear system:
given the external field $F$, the equation can be solved
for each value of $\omega$, having as unknowns the values of $\delta 
{\cal R}$ on the chosen basis. The solution of the linear system 
scales much more favourably than that of a matrix diagonalisation. 

\begin{figure}[t]
\includegraphics[width=0.55\textwidth]{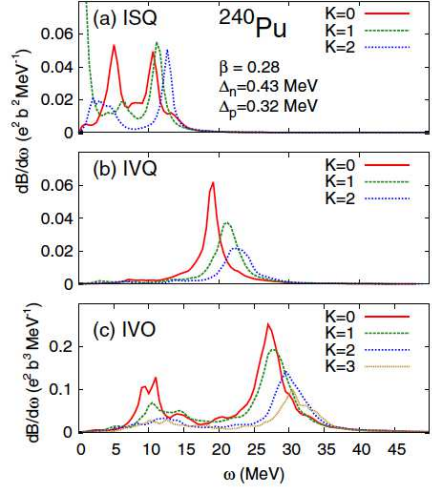}
\caption{Isoscalar (a) and isovector (b) quadrupole strengths, and 
isovector octupole strength (c) in the nucleus $^{240}$Pu,
calculated with FAM and a Skyrme functional. The figure is taken 
from Ref. (\cite{Kortelainen2015}), to which we refer for all details.}
\label{fig:7}
\end{figure}

In Fig. \ref{fig:7}, we show an example of FAM calculation. The 
multipole strength in a heavy system like $^{240}$Pu, with an
axial deformation associated with $\beta=0.28$, would be extremely
hard to calculate in a conventional matrix formulation. 
The formalism has been also implemented with relativistic functionals, 
and a computer code has been published in Ref. (\cite{Bjelcic2020}). 

\section{Reduction to RPA and the schematic model}

As we mentioned, in non-superfluid nuclei the HFB solution reduces to the simple
HF solution. The $u$ and $v$ coefficients are either 0 or 1, and the pair tensor vanishes.
In this condition, QRPA reduces to simple RPA. Eq. (\ref{eq:matrix}) keeps the same form
while the matrix elements $A$ and $B$ have a simpler form. They reduce to
\begin{eqnarray}\label{eq:rpa}
A_{ph,p'h'} & = & + \delta_{pp'}\delta_{hh'} \left( \varepsilon_p - \varepsilon_h \right)+ 
\langle ph' \vert V \vert hp' \rangle, \\
B_{ph,p'h'} & = & \langle pp' \vert V \vert hh' \rangle,
\label{eq:rpa2}
\end{eqnarray}
where $\varepsilon$ are the single-particle energies that we have introduced
in (\ref{eq:HF}). The matrix elements that enter the strength function 
(\ref{eq:strength}) can be calculated from 
\begin{equation}\label{eq:strength_rpa}
\langle n \vert F \vert 0 \rangle = \sum_{ph} \left( X_{ph} 
+ Y_{ph} \right)
\langle p \vert F \vert h \rangle.
\end{equation}
As in the case of QRPA, the reader can find details and formulas written
with proper angular momentum coupling in the general references that have
been quoted above. 

The $B$ matrix elements, and the associated $Y$ amplitudes are very important 
to describe the low-lying excitations like 2$^+$ and 3$^-$ in spherical nuclei, but turn
out to be less important for high-energy states like the GRs. Neglecting the $B$-sector
of the RPA matrix leads to the so-called Tamm-Dancoff approximation or TDA. 
We will now see how a schematic TDA model accounts for the existence of collective
modes like the IVGDR, in a pedagogical and transparent manner. 

\begin{figure}[t]
\includegraphics[width=0.95\textwidth]{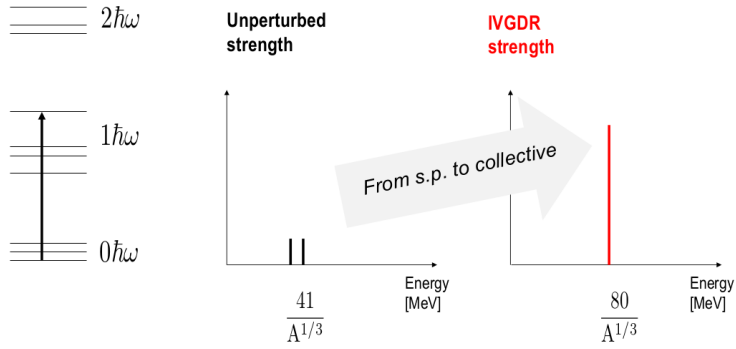}
\vspace{-1.0cm}
\caption{The schematic TDA model for the IVGDR. In the left panel, the nuclear 
shells are displayed in order to emphasise their alternate parity and the fact that 
we expect the IVGDR to be made by particle-hole excitations at $\approx$ 1$\hslash\omega$. 
In the central and right panels, we draw the unperturbed and TDA strength $S(E)$, 
respectively. See the text for a discussion.}
\label{fig:2}
\end{figure}

To this aim, let us assume that we have $N$ degenerate p-h excitations at
energy $\varepsilon$. In the case of the IVGDR, which has $J^\pi=1^-$, this corresponds to a
large extent to holes in the highest occupied shell and particles in the lowest 
unoccupied shell. Consequently, $\varepsilon$ will be $\approx 1\hslash\omega \approx 
41\ {\rm A}^{-1/3}$, as emphasised by the vertical arrow 
in the left panel of Fig. \ref{fig:2}.

We also consider the matrix elements that appear in $A$ as being all equal to a 
constant $v$. The TDA matrix becomes:
\begin{equation}\label{eq:schemTDA}
\left( \begin{array}{cccc} \varepsilon + v & v & \ldots & v \\
v & \varepsilon + v & \ldots & v \\ \ldots & & & \ldots \\
v & v & \ldots & \varepsilon + v \end{array} \right).
\end{equation}
Such simple model is exactly solvable. In the two by two case the matrix reads
\begin{equation}
\left( \begin{array}{cc} \varepsilon + v & v \\ v & \varepsilon + v \end{array} \right),
\end{equation}
and the eigenvalues and eigenvectors are
\begin{eqnarray}
\hslash\omega_1 & = & \varepsilon, \ \ \ \ \ X^{(1)} = \frac{1}{\sqrt{2}}\left( \begin{array}{c}
1 \\ -1 \end{array} \right), \nonumber \\
\hslash\omega_2 & = & \varepsilon+v \ \ \ \ \ X^{(2)} = \frac{1}{\sqrt{2}}\left( \begin{array}{c}
1 \\ 1 \end{array} \right).
\end{eqnarray}
We may call ``incoherent'' the first state because the two amplitudes have opposite phase, 
whereas the second state may be said to be ``coherent''. This wording will become clearer
if we calculate the matrix elements defined by Eq. (\ref{eq:strength_rpa}):
\begin{equation}\label{eq:me_F_TDA}
\langle n \vert F \vert 0 \rangle = \sum_{ph} X_{ph} \langle p \vert F \vert h 
\rangle.
\end{equation}
Consistently with the assumptions that have been already made, we can approximate the matrix 
elements $\langle p \vert F \vert h \rangle$ with a constant number $M$. Then, if one 
takes the incoherent state, the transition amplitude (\ref{eq:me_F_TDA}) associated with 
$F$ vanishes. Instead, its value is increased by a factor $\sqrt{2}$ with respect to the 
``single-particle value'' $M$ in the coherent case.

The schematic TDA equation (\ref{eq:schemTDA}) can also be solved in the general case with 
$N$\ $ph$ transitions. Also in that case, one finds there is only one coherent state, 
characterised by an eigenvalue $\varepsilon + Nv$ and by a corresponding eigenvector of the type
\begin{equation}
\frac{1}{\sqrt{N}}\left( \begin{array}{c} 1 \\ 1 \\ \ldots \\ 1 \end{array} \right).
\end{equation}
The transition amplitude (\ref{eq:me_F_TDA}) becomes
\begin{equation}
\langle n \vert F \vert 0 \rangle = \sum_{ph} X_{ph} \langle p \vert F \vert h
\rangle \approx N \frac{1}{\sqrt{N}} M = \sqrt{N}\ M.
\end{equation}
The quantity that appears in the strength function (\ref{eq:strength}) is the square 
of the transition amplitude, that is
\begin{equation}\label{eq:probcoll}
\vert \langle n \vert F \vert 0 \rangle \vert^2 = N\ M^2.
\end{equation}
In the case of the IVGDR, we clearly see that the term $Nv$ shifts the coherent eigenvalue 
from the unperturbed value $ \varepsilon \approx 41\ {\rm A}^{-1/3}$ to the value
$80\ {\rm A}^{-1/3}$ already mentioned in Eq. (\ref{eq:eneIVGDR}). The fact that this is
a ``giant'', or collective, state is highlighted by Eq. (\ref{eq:probcoll}): its probability to be
excited by the external field becomes $N$ times larger than the typical probability 
$M^2$ of a single-particle state. The situation is depicted in Fig. \ref{fig:2}.
In the central panel, we show the hypotetical single-particle, or unperturbed, 
strength at energy $\varepsilon$, which is made by several peaks whose height 
is $M^2$. In the right panel, we display instead the Giant Resonance made by a 
single peak whose height is larger by the factor $N$. Schematic models have been 
widely introduced throughout the years to describe collective states, and not 
only the IVGDR, starting from the seminal work of Ref. (\cite{Brown1959}). 

\section{Operators and example of calculations}

The GRs are excited by external operators that are classified 
as follows (\cite{Harakeh2001}). In the non spin-flip
case (electric GRs), the operators $F$ associated with different values of
the transferred angular momentum $L$ read
\begin{eqnarray}
F_{\rm IS} & = & \sum_i r_i^L Y_{LM}(\hat r_i), \label{eq:ISGR} \\
F_{\rm IV} & = & \sum_i r_i^L Y_{LM}(\hat r_i)\tau_z(i). \label{eq:IVGR}
\end{eqnarray}
In this way, we cleary distinguish isoscalar (IS) from isovector (IV) modes, that is,
cases in which protons and neutrons oscillate in phase from cases in which they move
against each other. $Y_{LM}$ are the usual spherical harmonics and $\hat r$ is, here and in what follows, 
a shorthand notation for the polar angles of $\vec r$. Values of $L=0, 1, 2, 3$\ldots 
correspond to monopole, dipole, quadrupole, octupole \ldots resonances. 
It is important to notice that the radial factor $r^L$ must be thought
as the limit of the Bessel function $j_L(qr)$ when $qr$ goes to zero (cf. the 
discussion in the next Section). When this term becomes meaningless, one must
consider the next term in the Taylor expansion of the Bessel function, that is
$r^{L+2}$. For instance, in the monopole case, $Y_{00}$ is a constant that 
can be neglected and the operator becomes
\begin{eqnarray}
F_{\rm ISGMR} & = & \sum_i r_i^2, \label{eq:ISGMR} \\
F_{\rm IVGMR} & = & \sum_i r_i^2 \tau_z(i). \label{eq:IVGMR}
\end{eqnarray}
The other case in which the term $r^L$ does not act as an excitation operator is the
isoscalar dipole case. $r Y_{1M}(\hat r)$ can be shown to produce a
translation of the whole nucleus, and also in this case the relevant operator becomes
proportional to $r^{L+2}$, namely
\begin{equation}
F_{\rm ISGDR} = \sum_i r_i^3 Y_{1M}(\hat r_i).  
\end{equation}
The nuclear translation, i.e. the motion of the nuclear centre of mass, should 
appear at zero excitation energy in an exact calculation, and should be decoupled
from the physical excitation modes. However, if this decoupling is not exactly
realised, a way to avoid the contamination of the physical ISGDR strength by 
the spurious translational strength, consists in employing the modified 
operator (\cite{VanGiai1981})
\begin{equation}
F_{\rm ISGDR} = \sum_i \left( r_i^3 - \eta r_i \right) Y_{1M}(\hat r_i), 
\end{equation}
where $\eta=\frac{5}{3}\langle r^2 \rangle$. 

In addition to isoscalar and isovector operators, one could be interested in the
electromagnetic excitation processes. In this case, the excitation operators become 
\begin{equation}
F_{\rm e.m.} = e \sum_i r_i^L Y_{LM}(\hat r_i) \left( 1 - \tau_z(i) \right).
\end{equation}
In the dipole case, it is customary to remove the contribution from the centre of 
mass, so that the IVGDR electromagnetic operator becomes
\begin{equation}
F_{\rm IVGDR} = \frac{eN}{A} \sum_{i=1}^Z r_i Y_{1M}(\hat r_i) 
- \frac{eZ}{A} \sum_{i=1}^N r_i Y_{1M}(\hat r_i).
\end{equation}
The effective charge for protons (neutrons) turns out to be $\frac{eN}{A}$ 
$\left( -\frac{eZ}{A} \right)$. In principle, this subtraction of the centre of mass 
should be done for every multipole. However, for $L \ge 2$, the resulting effective 
charges do not differ significantly from 1 and 0 [see their expression,
e.g. on p. 98 of Ref. (\cite{Eisenberg1976})].

In the spin-flip case (magnetic GRs) we write, in a similar way as in 
(\ref{eq:ISGR}) and (\ref{eq:IVGR}), 
\begin{eqnarray}
F_{\rm IS} & = & \sum_i r_i^L \left[ Y_{LM}(\hat r_i) \otimes \vec\sigma(i) \right]_J, 
\label{eq:spinIS} \\
F_{\rm IV} & = & \sum_i r_i^L \left[ Y_{LM}(\hat r_i) \otimes \vec\sigma(i) \right]_J
\tau_z(i). \label{eq:spinIV}
\end{eqnarray}
As we said, $J^\pi$ are good quantum numbers; but, as far as the operators 
are concerned, we are considering also $L$ and $S$ as approximate quantum numbers. 

The operators (\ref{eq:IVGR}) and (\ref{eq:spinIV}) correspond to
transitions within the same nucleus. There exist charge-exchange GRs, 
that correspond to the case in which the operator $\tau_z$ is replaced 
by $\tau_\pm$. The excited states of a given nucleus are, then, 
in the neighbouring $Z\mp 1$ isotopes: these are states that can be populated 
by $\beta$-decay, if energetically possible, plus those at higher energy. 
In what follows, we will show examples of strength functions in the case
of the Gamow-Teller Resonance (GTR), that is for the operator 
(\ref{eq:spinIV}) in the case $L=0$,
\begin{equation}
F_{\rm GTR} = \sum_i \sigma_\mu(i) \tau_-(i).
\end{equation}

\begin{figure}[t]
\centering\includegraphics[height=0.35\textheight]{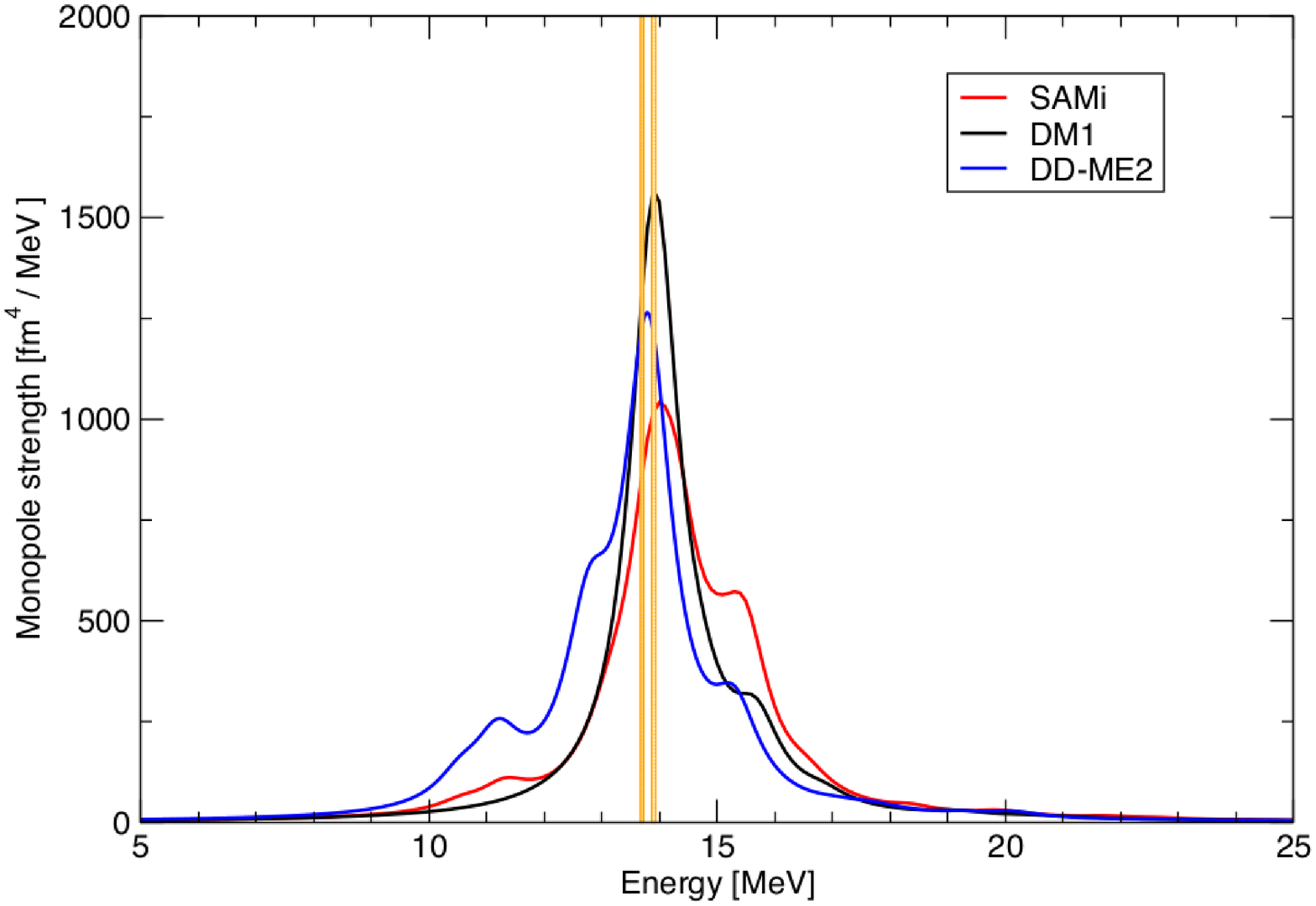}
\caption{RPA calculations of the monopole strength in $^{208}$Pb,
performed by using the nonrelativistic Skyrme-type functional
SAMi (\cite{Roca-Maza2012}), the Gogny-type functional DM1
(\cite{Goriely2009}), and the relativistic functional DDME2 
(\cite{Lalazissis2005}). The discrete RPA peaks are smeared out with
Lorentzian functions having a width of 1 MeV [Eq. (\ref{eq:strength_gamma}) 
is used with a fixed width]. The vertical lines show the peak 
energy obtained in the experiments performed at TAMU (13.9 MeV) 
and RCNP (13.7 MeV). Figure taken from (\cite{Garg2018}), where 
the original references for the experimental data can be found. 
}
\label{fig:3}
\end{figure}

We display in this Section some realistic examples of RPA and QRPA 
calculations. 
In Fig. \ref{fig:3}, one can see the results of RPA calculations of 
the ISGMR in the spherical nucleus $^{208}$Pb, performed with different EDFs, of Skyrme, 
Gogny and covariant type. 
The figure is meant to illustrate the fact that RPA can reproduce well the
experimental value of the ISGMR peak energy, provided the Hamiltonian or EDF is well
calibrated. In this case, the different EDFs are all associated with a realistic value
of the nuclear incompressibility $K_\infty$, and the result does not depend much on
whether a Skyrme, Gogny or covariant functional is used.

\begin{figure}[t] 
\centering\includegraphics[width=12.0cm]{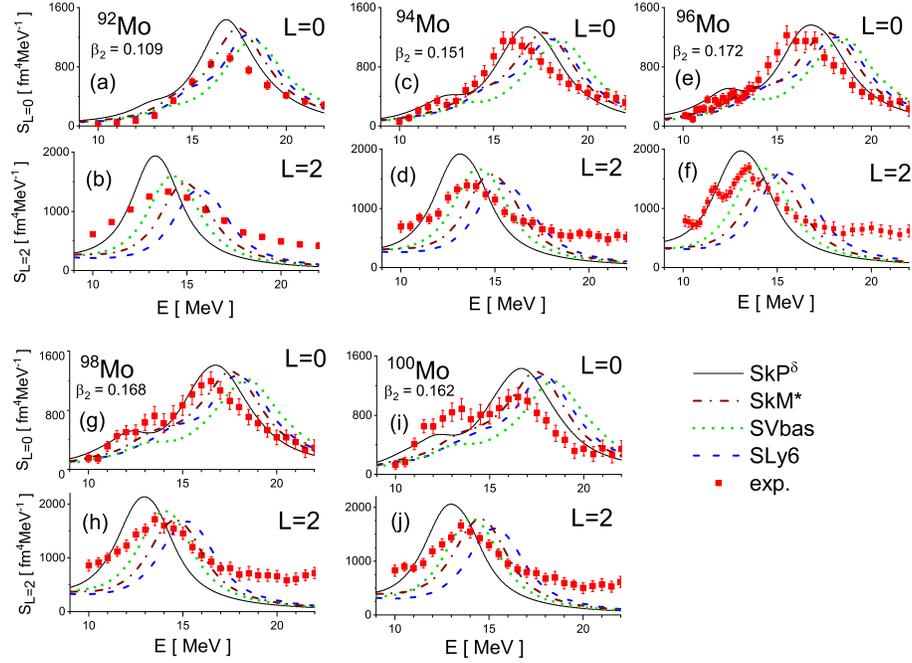}
\caption{Isoscalar monopole and quadrupole strengths in $^{92,94,96,98,100}$Mo, calculated
within QRPA and using the Skyrme forces SkM$^*$, SLy6, SVbas and SkP$^{\delta}$. The strengths
are compared with the experimental data. The figure is taken from (\cite{Colo2020}), where
references to the experimental data and details on the method used can be found.}
\label{fig:4}
\end{figure}

In Fig. \ref{fig:4}, we show QRPA calculations of the ISGMR
that have been performed without assuming spherical symmetry along the chain of the
Mo isotopes. These isotopes display a modest deformation which is enough, nonetheless, 
to produce an observable effect. In fact, for axially deformed nuclei only the projection
of $J$ along the symmetry axis, called $K$, is a good quantum number. The calculations
are performed in a space with good $K^\pi$, so that the monopole states are coupled to the $K=0$ 
component of the quadrupole states (in principle, they are coupled also with higher
multipoles but this coupling becomes significantly smaller with increasing angular momentum). 
The monopole-quadrupole coupling is evident from Fig. \ref{fig:4}. The use of
operators of the type $F = r^L Y_L$ helps to partially decouple the multipoles 
(\cite{Arteaga2008}), but a full angular momentum projection would be called for.
Similar calculations have been performed by other groups [cf. 
(\cite{Yoshida2021}) and references therein].

\section{Relationship between strength functions and inelastic cross sections}

State-of-the-art reaction calculations are performed within the so-called Distorted-Wave
Born Approximation (DWBA), in which the differential cross section is given by
\begin{equation}\label{eq:DWBA}
\frac{d\sigma}{d\Omega} = 
\frac{k_f}{k_i} \vert f(\theta) \vert^2 =
\frac{\mu^2 k_f}{4\pi^2\hslash^4k_i}
\left\vert \int d^3R\ \chi^*_{{\bf k}_f}({\bf R})\
\langle f \vert V \vert i \rangle\ \chi_{{\bf k}_i}({\bf R}) \right\vert^2.
\end{equation}
Here, $V$ is a projectile-target interaction and $\chi$ are the distorted waves in the
initial and final channels, in which the relative motion is associated 
with the momenta $k_i$ and
$k_f$, respectively. In what follows, we will assume $k_f \approx k_i$. 
$\bf R$ is the relative projectile-target coordinate and $\mu$ is the
reduced mass. Although in inelastic scattering the mutual interaction $V$ can act as a field
to excite one of the two nuclei, Eq. (\ref{eq:DWBA}) does not display any simple
relationship with the strength function that we have discussed so far.

To unveil this relationship, we are obliged to drastically simplify our formulation. We 
resort to the Plane-Wave Born Approximation (PWBA) and we assume that the interaction $V$ 
is a zero-range force. Within the PWBA, if ${\bf q}$ is the momentum 
transfer, the scattering amplitude $f(\theta)$ is simply
\begin{eqnarray}\label{eq:fPWBA}
f(\theta) & = & -\frac{\mu}{2\pi\hslash^2}\int d^3R\ e^{i{\vec q}\cdot{\vec R}}
\langle f \vert V \vert i \rangle \nonumber \\
& = & -\frac{2\mu}{\hslash^2}\sum_{LM} i^L\ \int d^3R\ j_L(qR) 
Y_{LM}(\hat R)Y_{LM}^*(\hat q)
\langle f \vert V \vert i \rangle,
\end{eqnarray}
where we have used the expansion of the plane wave in spherical components. Accordingly, 
we also expand in multipoles the interaction $V$, and we neglect its spin- and 
isospin-dependence for the sake of simplicity:
\begin{equation}\label{eq:deltaforce}
V \equiv \sum_i V({\bf R} - {\bf r}_i) = V_0 \sum_i \delta({\bf R} - {\bf r}_i) =
V_0 \sum_i \sum_{\lambda\mu}\frac{\delta(R-r_i)}{R^2}Y_{\lambda\mu}(\hat r_i)Y^*_{\lambda\mu}(\hat R).
\end{equation}
We insert Eq. (\ref{eq:deltaforce}) in (\ref{eq:fPWBA}) and we obtain
\begin{equation}\label{eq:f_final}
f(\theta) = -\frac{2\mu V_0}{\hslash^2}\sum_{LM} i^L Y^*_{LM}(\hat q)
\sum_i \langle f \vert j_L(qr_i) Y_{LM}(\hat r_i) \vert i \rangle.
\end{equation}
In case of experiments with unpolarised beams, in order to obtain the cross section, we have to average 
over the possible values of $M$ of the initial state and sum 
over the values of $M$ of the final state. 
This is simple in the case of $L=0$ initial state, where the final 
$M$-value and the transferred $M$ of Eq. (\ref{eq:f_final}) coincide. 
Eventually, for a given 
$L$-transfer, 
\begin{equation}\label{eq:final_sigma}
\frac{d\sigma}{d\Omega} = \left( \frac{2\mu V_0}{\hslash^2} \right)^2 
\frac{
\vert \langle f \vert\vert j_L Y_L \vert\vert i \rangle \vert^2}
{4\pi}\ P_L(cos(\theta)),
\end{equation}
where $P_L$ is a Legendre polynomial. 
If the momentum transfer $q$ is small enough 
so that the condition $qR \ll 1$ holds, we can exploit the fact that the Bessel functions 
behave as $j_L(qR)\approx (qR)^L$. In this case, one can see clearly
that the cross section (\ref{eq:final_sigma}) factorises into a kinematical
part and the square of the matrix element of the operators that we have
already displayed in Eq. (\ref{eq:ISGR}), that are those commonly used
in RPA and its extensions. 

We stress, once again, that this decomposition into a kinematical factor and
the square of the matrix element of a radial multipole operator is valid 
only within the plane-wave approximation. In reality, the 
cross section will be ``distorted'' with respect to the strength function. 
However, the relationship that we
have highlighted possesses its conceptual validity as a guideline.

\section{More general frameworks beyond (Q)RPA}

The natural extension of QRPA is second QRPA, in which the excited states are supposed
to have not only two quasi-particle components but also four quasi-particles components,
\boxedeqnlarge{
\vert N \rangle = \left( \sum_{\alpha\beta} 
X^{(1)}_{\alpha\beta} \alpha^\dagger_\alpha\alpha^\dagger_\beta 
- Y^{(1)}_{\alpha\beta}\alpha_\beta\alpha_\alpha +
\sum_{\alpha\beta\gamma\delta} X^{(2)}_{\alpha\beta\gamma\delta} 
\alpha^\dagger_\alpha\alpha^\dagger_\beta\alpha^\dagger_\gamma\alpha^\dagger_\delta
- Y^{(2)}_{\alpha\beta\gamma\delta} 
\alpha_\delta\alpha_\gamma\alpha_\beta\alpha_\alpha \right) \vert \tilde 0 \rangle.
\label{eq:SRPA_basis}
}
In the non-superfluid
case second QRPA reduces to second RPA (SRPA), and the states $\alpha\beta$ and
$\alpha\beta\gamma\delta$ are replaced by $ph$ and $php'h'$, respectively. With a procedure 
similar to that outlined in our previous discussion, the SRPA equations can be obtained by 
demanding that the eigenstates of the system have the more general form 
(\ref{eq:SRPA_basis}) instead of the simpler form (\ref{eq:QRPAstates}).  
Their derivation can be found in textbooks like (\cite{Rowe1970}) [cf. also 
(\cite{Yannouleas1987}) and the more recent (\cite{Papakonstantinou2014}) 
for their formal properties]. One shoud note
that the SRPA ground-state is still more general than the ground-state
$\vert \tilde 0\rangle$ that we have discussed in the paragraph above
Eq. (\ref{eq:QRPAstates}); nevertheless, we have avoided to introduce
a more complicated notation. 

The SRPA equations can be cast in a matrix form, that is,
\begin{equation}\label{eq:fullSRPA}
\left( \begin{array}{cccc} A_{11}& A_{12} & B_{11} & B_{12} \\ 
A_{21}& A_{22} & B_{21} & B_{22} \\ 
-B_{11}^* & -B_{12}^* & -A_{11}^* & -A_{12}^* \\
-B_{21}^* & -B_{22}^* & -A_{21}^* & -A_{22}^* 
\end{array} \right) \left( \begin{array}{c} 
X^{(1)} \\ X^{(2)} \\ Y^{(1)} \\ Y^{(2)} \end{array} \right) = \hslash\omega 
\left( \begin{array}{c} X^{(1)} \\  X^{(2)} \\ Y^{(1)} \\ Y^{(2)} \end{array} \right).
\end{equation}
In fact, this equation is analogous to (\ref{eq:matrix}), except for the fact that the matrices
$A$ and $B$ have larger dimension. The indices $1$ and $2$ refer to the part of the
model space that includes 1p-1h and 2p-2h configurations, respectively. As discussed
in (\cite{Yannouleas1987}), the expressions of $A$ and $B$ can be obtained after a 
lengthy but straightforward calculation. The detailed formulas are not reported here
but we stress that $A$ and $B_{11}$ do not vanish, whereas $B_{21} = B_{12} = B_{22} = 0$.

If we use standard projection techniques to reduce (\ref{eq:fullSRPA}) to an equilvalent
equation in the smaller 1p-1h space, we find
\begin{equation}\label{eq:projected_matrix}
\left( \begin{array}{cc} {\cal A}(\omega) & {\cal B} \\ -{\cal B}^* & -{\cal A}^*(-\omega)
\end{array} \right) \left( \begin{array}{c} {\cal X} \\
{\cal Y} \end{array} \right) = \hslash\omega \left( \begin{array}{c} {\cal X} \\
{\cal Y} \end{array} \right), 
\end{equation}
where 
\begin{eqnarray}\label{eq:SRPA_me}
{\cal A}(\omega) & = & A_{11} + \Sigma(\omega), \nonumber \\
{\cal B} & = & B_{11},
\end{eqnarray}
and $\Sigma$ is the self-energy of the 1p-1h configurations due to the coupling with 2p-2h,
\begin{equation}\label{eq:sigma1}
\Sigma(\omega) = A_{12}\frac{1}{\hslash\omega - A_{22}}A_{21}.
\end{equation}

Although SRPA has been proposed as a theory a long time ago, during several decades calculations
were too demanding especially for medium-heavy nuclei. One of the most common approximations
that have been adopted, consisted in assuming the 2p-2h states as non interacting. In this way,
the matrix $A_{22}$ is diagonal and the computational effort is not much heavier than RPA 
[Eq. (\ref{eq:sigma1}) contains only a sum]. 
This is called indeed ``diagonal approximation''. 
More recently, realistic SRPA calculations that
lift this severe approximation have been performed. 

SRPA can be presently performed even with realistic interactions, as
demonstrated in (\cite{Papakonstantinou2010}). The quality of the results, 
in terms of favourable comparison with experiment, depends on the 
adopted Hamiltonian. In fact, we should mention an interesting key open
question at this point. The self-energy (\ref{eq:sigma1}) is an
increasing function of $\omega$, that tends to shift a significant 
amount of strength downward in energy. In the case of a zero-range
interaction like the Skyrme one, the self-energy does actually diverge
if $\omega$ becomes arbitrarily large. But even with finite-range
iteractions like Gogny, the accumulation of low-energy strength 
that was found in SRPA calculations [see, e.g., (\cite{Gambacurta2015})] was 
found to be unrealistically large and in contrast with the experimental findings. 
This may be related to the Hamiltonian or EDF that is employed and/or may be
reduced by introducing further correlations beyond SRPA.

\begin{figure}[t]
\includegraphics[width=0.65\textwidth]{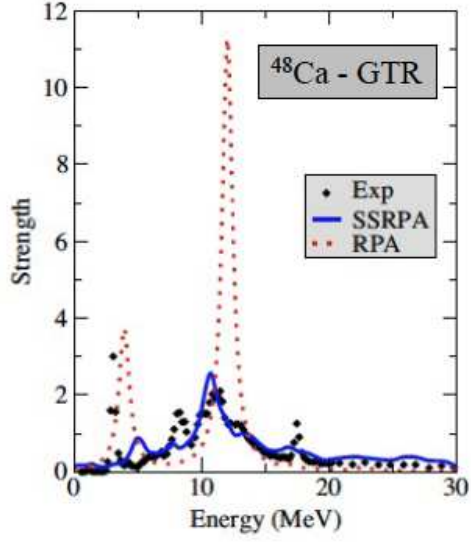}
\caption{Gamow-Teller strength in $^{48}$Ca calculated in RPA and SRPA with subtraction 
(SSRPA), using the Skyrme interaction SGII. Figure taken from (\cite{Gambacurta2020}), 
where details and the reference to the experimental data can be found.}
\label{fig:5}
\end{figure}

Definitely, effective Hamiltonians or EDFs are fitted at mean-field level. When
practitioners check if the Hamiltonian of EDF is ``well calibrated'' for 
the GR sector (as we mentioned when discussing that EDFs provide a
correct ISGMR energy if associated with the correct value of $K_\infty$), they do it at
the level of RPA. SRPA (as is also true for the models to be introduced below) produces
an additional shift of the 
states $\vert N \rangle$ of (\ref{eq:SRPA_basis}) 
towards lower energies, in addition to broadening the strength 
and making it more fragmented because the states $\vert N \rangle$ are many more
than the RPA states.
In Ref. (\cite{Tselyaev2013}), it has been argued that, in order to include effects beyond 
mean-field without altering the properties that a sensible EDF must keep, one has to
implement SRPA using the so-called ``subtraction technique''. This means replacing, in 
Eq. (\ref{eq:SRPA_me}),
\begin{equation}
\Sigma(\omega) \ \rightarrow \ \Sigma(\omega)-\Sigma(\omega=0).
\end{equation}
This prescription avoids also possible divergences of the self-energy 
(to be definitely expected in the case of zero-range forces) 
when the upper limit
on the energy of the 2p-2h states is increased. On the other hand, 
its grounds in a general many-body theory, as well as its 
implications for other studies like the shift and fragmentation
of the single-particle strength, have not been yet fully clarified. We 
show in
Fig. \ref{fig:5} an example of a recent calculation using the subtraction
technique, labelled as subtracted-SRPA or SSRPA.

In (Q)RPA plus particle-vibration coupling [(Q)RPA+PVC], one adpots a similar philosophy as
in SRPA but the four quasi-particle (or 2p-2h) states are replaced by two quasi-particles
(or 1p-1h) plus a vibration or phonon. This means that Eq. (\ref{eq:SRPA_basis}) becomes
\boxedeqn{\label{eq:PVC_basis}
\vert N \rangle = \left( \sum_{\alpha\beta}
X^{(1)}_{\alpha\beta} \alpha^\dagger_\alpha\alpha^\dagger_\beta
- Y^{(1)}_{\alpha\beta}\alpha_\beta\alpha_\alpha
+ 
\sum_{\alpha\beta n} X^{(2)}_{\alpha\beta n}
\alpha^\dagger_\alpha\alpha^\dagger_\beta\Gamma^\dagger_n
- Y^{(2)}_{\alpha\beta n}
\Gamma_n\alpha_\beta\alpha_\alpha\right) \vert \tilde 0 \rangle,
}
where $\Gamma_n^\dagger$ is the creator of the QRPA state $\vert n\rangle$ in
(\ref{eq:QRPAstates}), that is
\begin{equation}\label{eq:Gamma_dagger}
\Gamma^\dagger_n = 
\sum_{\alpha\beta} X^{(n)}_{\alpha\beta}
\alpha^\dagger_\alpha\alpha^\dagger_\beta - Y^{(n)}_{\alpha\beta}
\alpha_\beta\alpha_\alpha.
\end{equation}
The rationale behind this choice is that, by replacing a pair of quasi-particles
with one vibration, one includes further correlations. 

More specifically, the vibrational states $\vert n\rangle$ lie, on
average, at lower energy than the two quasi-particle states. In other
words, the states labelled by $\alpha\beta n$ in (\ref{eq:PVC_basis}) 
lie, on average, at lower energy than the states labelled by 
$\alpha\beta\gamma\delta$ in (\ref{eq:SRPA_basis}). Either class of
states is commonly referred to with the wording ``doorway states'': they
constitute the ``doorway'' to GR damping. The doorway states that lie
at lower energies are more likely to be coupled to the dominant 
two quasi-particle components: therefore, (Q)RPA+PVC is more effective in
reproducing the GR widths than SRPA. On the other hand, the doorway
states that appear in (Q)RPA+PVC form an overcomplete basis and violate,
to some extent, the Pauli principle while the same does not happen in SRPA.
We will discuss further this issue here below.

The first implementation of this model using Skyrme interactions has been described in detail in 
(\cite{Colo1994}). The same model has been resumed in (\cite{Roca-Maza2017}) for
non charge-exchange states and in (\cite{Niu2014}) for charge-exchange states.
All these calculations can be understood along the same line of the discussion 
made for SRPA. A matrix which is fully comparable to  
(\ref{eq:projected_matrix}) is diagonalised, and the self-energy is also
analogous to that of SRPA in the diagonal approximation. Eq. 
(\ref{eq:sigma1}) holds, but it includes the coupling of 1p-1h to
1p-1h plus a phonon $n$. By writing it in detail, we obtain
\begin{equation}\label{eq:sigma2}
\Sigma_{ph,p'h'} (\omega) = \sum_{p^{\prime\prime}h^{\prime\prime}n} 
A_{ph,p^{\prime\prime}h^{\prime\prime}n}
\frac{1}{\hslash\omega - \left( \varepsilon_{p^{\prime\prime}} 
- \varepsilon_{h^{\prime\prime}} + \hslash\omega_n \right) } 
A_{p^{\prime\prime}h^{\prime\prime}n,p'h'}.
\end{equation}
In Ref. (\cite{Shen2020}), the approximation of completely non-interacting 1p-1h-1 phonon states
has been removed. All these works deal with the implementation without pairing. QRPA+PVC 
for non charge-exchange states has been introduced, using Skyrme Hamiltonians 
and based on HF-BCS, in Ref. (\cite{Colo2001}). More recently, 
QRPA+PVC for charge-exchange states on the HFB basis has been introduced in
(\cite{Niu2016}). In all these works, an overall satisfactory description
of the lineshape of GRs, namely of their width and fragmentation, has been
achieved.

\begin{figure}[t]
\includegraphics[width=0.95\textwidth]{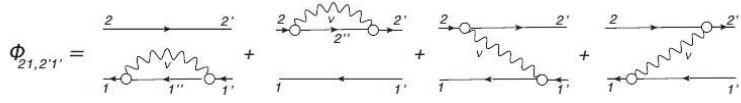}
\caption{The so-called ``dynamical kernel'' of the Time-Blocking 
Approximation (TBA). See the text for a short discussion. Figure taken from (\cite{Litvinova2019}).}
\label{fig:8}
\end{figure}

There is quite a strict relationship between (Q)RPA+PVC and the Time-Blocking
Approximation (TBA). This theory has been introduced in the context of
phenomenological calculations based on a Woods-Saxon mean-field plus a
residual Landau-Migdal interaction (\cite{Kamerdzhiev2004}). Nevertheless, 
there exists a series of recent GR calculations using Skyrme functionals
[cf. \cite{Tselyaev2016} and references therein]. In the basic version,
TBA is actually the same framework as the one we have just described under the 
name of (Q)RPA+PVC. In Fig. \ref{fig:8}, we depict the Feynman diagrams 
that correspond to the so-called ``dynamical kernel'' of this theory. These
diagrams correspond, in fact, to the coupling between states 12 and 1'2' 
belonging to the two quasi-particle space with intermediate doorway states
given by 1p-1h plus a phonon $\nu$. Thus, the quantity $\Phi$ in
Fig. \ref{fig:8} is exactly the self-energy $\Sigma(\omega)$ of
Eq. (\ref{eq:sigma2}). In the calculations reported in Ref. 
(\cite{Tselyaev2016}), the coupling to the continuum is to some extent
included [as it was in (\cite{Colo1994})]. The description of the width
of the GRs in magic, medium-heavy nuclei is rather good, although this is less true in
light nuclei as $^{16}$O. Whether this is due to missing decay channels
like the $\alpha$-particle decay, or to other limitations of the model,
is still unclear.

\begin{figure}[t]
\includegraphics[width=0.85\textwidth]{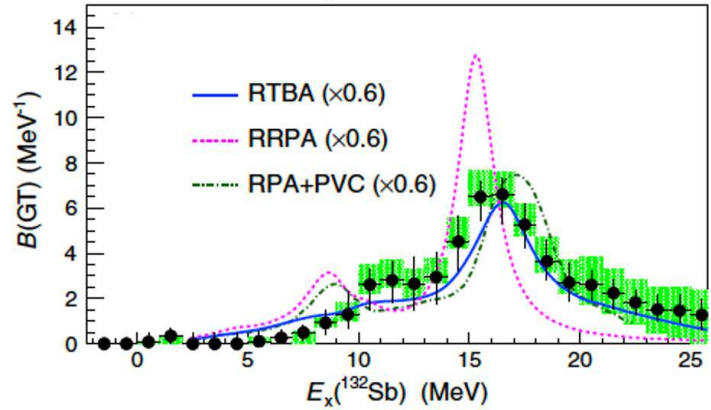}
\caption{Gamow-Teller strength in $^{132}$Sn. The experimental data and the figure 
are taken from (\cite{Yasuda2018}), where more details can be
found. The figure shows a comparison of the performance
of different theoretical approaches discussed in this chapter.}
\label{fig:6}
\end{figure}

In Fig. \ref{fig:6}, we show results for a case in which we can compare 
simple relativistic RPA (RRPA), in which a small width is put by hand, 
with relativistic TBA (RTBA) and RPA+PVC. The differences between these
last two approaches should be ascribed to the different functionals
employed. RTBA has been introduced in Refs. (\cite{Litvinova2007}) 
and (\cite{Litvinova2008}) for magic and superfluid nuclei, respectively.
Recently, progress has been made in pushing this theory beyond
the level we have described in this Section. In the work of Ref. 
(\cite{Robin2019}), on top of the diagrams already depicted in Fig. 
\ref{fig:8}, further ones of second-order in the particle-phonon
interaction have been introduced in the self-energy $\Sigma(\omega)$. 
Even more interestingly, RTBA has been recast as a specific approximation
within the general many-body theory in Ref. (\cite{Litvinova2019}) [cf. 
also Ref. (\cite{Litvinova2020}) for the superfluid case].

In all the models based on the interplay of particles and vibrations, 
or phonons, one of the issues is the number of phonons included and 
the convergence of the results with respect to this input. In this 
respect, the convergence of the GR peak energy with respect to 
the model space has been carefully studied in (\cite{Tselyaev2016}). 
The convergence with respect to the maximum phonon energy is well 
realised; on the other hand, regarding the phonon strength, 
the problem is that there is moderate dependence (optimistically,
a sort of plateau) when one decreases the cutoff strength of the 
phonon to about 10\%. It has to be noted that introducing phonons
with small values of strength, i.e. non collective, raises the issue
of the Pauli principle violations that we have already mentioned.
In principle, in all the theories that we have discussed, precise
prescriptions exist to restore the Pauli principle; in practice, 
however, exact prescriptions are seldom used and practitioners often employ
a practical recipe that consists in using only collective phonons.

We end this Section by mentioning another model that shares a similar
point of view. The Quasiparticle-Phonon Model (QPM) describes GRs 
as admixtures of one phonon components, two phonons components, etc. 
The single phonon is built using QRPA, that is, the phonon creation 
operator is given by Eq. (\ref{eq:Gamma_dagger}). The equation corresponding
to (\ref{eq:SRPA_basis}) or to (\ref{eq:PVC_basis}) reads
\boxedeqn{\label{eq:QPM_basis}
\vert N \rangle = \sum_n R_n \Gamma^\dagger_n + \sum_{nn'} P_{nn'} 
\Gamma^\dagger_n\Gamma^\dagger_{n'} \vert \tilde 0 \rangle,
} 
where $R_n$ and $P_{nn'}$ are unknown amplitudes to be determined.
The basic theory can be found in (\cite{Soloviev1992}). 
Within the QPM, detailed equations
to correct for the Pauli principle violations have been written and, to a large extent, 
implemented. On the other hand, QPM has been mostly implemented using 
phenomenological input (a Woods-Saxon mean-field plus schematic pairing 
and a separable residual interaction).
Among the countless applications of QPM to the analysis of GRs, we single out the studies
of double GRs [cf. (\cite{Ponomarev2000,Ponomarev2001}) 
and references therein], and the analysis
of the low-lying pygmy dipole strength and its IS/IV character (\cite{Savran2008,Endres2010}).

\section{Width, fine structure, particle- and $\gamma$-decay}

We have already discussed the width of GRs in the previous sections. 
The purpose of this specific and short section is to distinguish
more carefully the different contributions to that width, and 
discuss the capability
of the different approaches to describe each of them. Again, the topic 
is not new and, in addition to the general textbooks, we should quote
the review paper (\cite{Bertsch1983}). 

(Q)RPA and the related approaches can account for the fact that 
there is not a single GR peak but the strength is instead fragmented 
(cf. the different figures in this chapter and, in particular, Fig. 
\ref{fig:4}). This fragmentation is, at times, referred to as the so-called 
``Landau width''; but this wording is somehow ambiguous
and ought to be better avoided. The total width $\Gamma$ is, then, given by the
following terms:
\begin{equation}\label{eq:gammatotal}
\Gamma = \Gamma^\uparrow + \Gamma^\downarrow + \Gamma_\gamma.
\end{equation}

The so-called ``escape width'' $\Gamma^\uparrow$ is associated with
the coupling to the continuum, that is, to the emission
of particles. GRs are, as a rule, above the threshold for particle emission
and can emit most likely neutrons but also protons and, in some cases, 
heavier residues like $\alpha$-particles. Continuum-RPA (CRPA) is
the model in which one considers p-h excitations with the particle
in the continuum part of the spectrum, without the discretisation
that is implicit in equations like (\ref{eq:matrix}), (\ref{eq:rpa}) 
and (\ref{eq:rpa2}). At the turn of the 20th century, a series of works 
have implemented CRPA fully self-consistently using Skyrme functionals:
the purpose was the analysis of GRs and low-energy multipole strength
in exotic nuclei close to the drip line (\cite{Hamamoto1997a},
\cite{Hamamoto1997}, \cite{Hamamoto1998}). Later, CRPA has been
implemented also in the case of finite-range effective interactions
like Gogny, both for normal (\cite{DeDonno2011}) and charge-exchange
excitations (\cite{DeDonno2016}).

To disentangle the escape width from the other contributions in
(\ref{eq:gammatotal}), one has to measure the emitted particles.
Experimental measurements of particle-decay are described in 
textbooks (\cite{Harakeh2001}). Both phenomenological 
[see (\cite{Gorelik2021}) and numerous references therein] and 
fully microscopic approaches (\cite{Colo1994}) have been proposed 
to calculate the partial decay widths $\Gamma_c^\uparrow$ ($c$ labels
a specific decay channel and by definition $\Gamma^\uparrow = 
\sum_c \Gamma_c^\uparrow$).   

$\Gamma^\downarrow$ is the so-called ``spreading width''. While
$\Gamma^\uparrow$ is associated with the damping into external
degrees of freedom, $\Gamma^\downarrow$ is associated with 
the damping into internal degrees of freedom. In other words, the spreading 
width corresponds to the finite lifetime with which the simple
configurations, like two quasi-particles, decay into progressively
more complex degrees of freedom like four quasi-particles, six
quasi-particles etc. (In magic nuclei, we would speak about
1p-1h that decay into 2p-2h, 3p-3h etc.) Except in light
nuclei, $\Gamma^\downarrow$ is by far dominant (\cite{Bertsch1983}). 
All the models that we have described in the last Section
(SRPA, QRPA+PVC, TBA and beyond, QPM) are based on, and support, the
idea that most of the damping is accounted for by the coupling
at the level of 2p-2h. The effect of 3p-3h coupling on the GRs
is arguably small (\cite{Ponomarev1996}). 

Recently, a series of high-resolution experiments have been carried out,
with the goal of understanding the underlying structure of the 
GR line shape. The question can be asked, whether the GRs are
best described as single Lorentzians with a large width, or in terms
of more complicated structures. To this aim, the so-called wavelet
analysis has been carried out by looking both at experimental strength
functions or cross sections, and at the theoretical results 
[cf. (\cite{Shevchenko2004}) and references therein]. The conclusion,
so far, is that similar scales appear in experiment and theory 
although quantitative differences show up when one compares them 
in detail. The idea that coupling at the level of 2p-2h is the dominant
mechanism seems to be supported by this analysis in different cases, 
like dipole and quadrupole resonances. 

Finally, $\Gamma_\gamma$ in Eq. (\ref{eq:gammatotal}) is a very tiny
contribution, of the order of $10^{-3}-10^{-4}$ of the total width.
The direct $\gamma$-decay is, nevertheless, a further probe of
the microscopic structure of GRs, as the other observables that
we have discussed in this Section like the fine structure and the
particle emission. The $\gamma$-decay of GRs to the ground state can be
calculated within (Q)RPA. In principle, electromagnetic transitions
from the GR to another phonon state could be also calculated at the
(Q)RPA level, but this goes against the fact that (Q)RPA as a framework
includes only lowest order processes. Calculations at the level
of QPM and RPA+PVC have been reported, respectively, in Refs. 
(\cite{Voronov1990}) and (\cite{Lv2021}). 

\section{Other topics}

There are essentially three current lines of research in the domain of 
GR physics: (i) use well-established experimental data as a benchmark for
new theories; (ii) identify GRs in neutron-rich, eventually weakly-bound 
isotopes close to the drip lines; (iii) identify elusive modes that 
have not been seen so far. 

Concerning (i), we should mention that, while DFT is a sort of mature
theory, still there is much work to be done to improve {\em ab initio}
calculations. {\em Ab initio} approaches use some technique to solve the many-body problem that
is in principle exact, or systematically improvable, and can provide a reliable estimate of
the theoretical error of the predicted quantities. 
Starting from the basic theory of strong interactions, that is Quantum 
Chromo Dynamics (QCD), there is no unique way to derive a low-energy
effective Hamiltonian. This would not be a problem at all, if there
were different Hamiltonians that predict compatible values for the different observables.
So far, the different Hamiltonians are not accurate enough, and there are discrepancies
between the results of different Hamiltonians when used within a given many-body method.
In this respect, GR physics is a formidable and unique playground to
fix procedures that guarantee that those low-energy effective Hamiltonians
capture the most important nuclear correlations. 

The multipole response of unstable nuclei, and the possible occurrence 
of new exotic modes of excitation in weakly-bound nuclear systems, 
has been the subject of the review paper (\cite{Paar2007}). 
Many of the theoretical tools that we have described in this chapter
were introduced in that review paper, except for the new
developments that have taken place after 2007. The evolution 
of the low-energy dipole response in unstable neutron-rich nuclei 
is a topic that has been subject of strong interest, and is also 
subject of another chapter in this Handbook. There exist
isotopic chains along which some low-energy or ``pygmy'' resonances
seem to develop; but, often, the available emprirical information 
is not sufficient to determine the nature of these pygmy resonances. 
Monopole, quadrupole, or spin pygmy states have been proposed 
in the recent literature. 

As we said, one often looks for possible new, elusive modes. This
may be simply curiosity-driven, or may be related to specific goals.
However, one of the strongest reasons to study GRs is their link
with the nuclear Equation of State (EoS). The EoS is the 
function (not functional here) $E[\rho_n,\rho_p]$ in uniform
nuclear matter. The review paper (\cite{Roca-Maza2018a})
summarises the present constraints to the EoS, around 
the saturation density of symmetric matter, from nuclear structure 
calculations of ground and collective excited state properties 
of atomic nuclei. Not only the ISGMR is linked to the nuclear incompressibility,
but the dipole response is connected to the so-called symmetry energy
and its density dependence. The symmetry energy $S(\rho)$ is the energy
per particle to change protons into neutrons, starting from symmetric matter
at a given density. Knowing this sector of the EoS is of paramount
importance for our understanding of dilute, neutron-rich systems, or
of neutron stars.

Often, a nice correlation shows up between GR properties 
(energy of the ISGMR, or of the IVGDR) and some parameter of the EoS.
This can be seen in (\cite{Garg2018}) and in several figures 
of (\cite{Roca-Maza2018a}) (cf. also the Table 2 in this latter 
paper, and the references quoted therein). These correlations
appear, as a rule, when working at the level of (Q)RPA. As we have 
discussed, the main effect of coupling to phonons or 2p-2h 
is a shift downward of the GR peak energy. In some cases, 
this shift is small enough so that it can be neglected [cf. the monopole case
as discussed in (\cite{Garg2018})]. More often, this shift is of the order 
of $\approx$ 1 MeV or slightly more. It does not markedly depend on the 
chosen effective Hamiltonian or EDF. Unfortunately, it breaks the 
nice correlation that exists between (Q)RPA results and EoS parameters. 
Pragmatic approaches, that amount to assuming that the correlation 
that holds at the (Q)RPA level is simply shifted by the aforementioned
quantity of about 1 MeV, could be assumed (\cite{Niu2012}, 
\cite{Tselyaev2016}). Analysing in a deeper manner 
the link between the GR properties and
the EoS beyond (Q)RPA is still an open issue.


\bibliographystyle{aps-nameyear}      
\bibliography{GR}                     
\nocite{*}                            

\end{document}